\documentclass[
reprint,
superscriptaddress,
amsmath,amssymb,
aps,prd, 
nofootinbib
]{revtex4-2}

\usepackage[colorlinks=true,allcolors=blue]{hyperref}
\usepackage{graphicx}
\usepackage{orcidlink}

\newcommand{\re}{\mathrm{Re}}
\DeclareMathOperator*{\E}{\mathrm{E}}
\DeclareMathOperator*{\Var}{\mathrm{Var}}
\DeclareMathOperator*{\atan2}{\mathrm{arctan2}}

\begin{document}

\title{Spectral signatures of axionlike dark matter}

\author{Alexander~V.~Gramolin\,\orcidlink{0000-0001-5436-7375}}
\affiliation{Department of Physics, Boston University, Boston, Massachusetts 02215, USA}

\author{Arne~Wickenbrock\,\orcidlink{0000-0001-5540-7519}}
\affiliation{Helmholtz-Institut, GSI Helmholtzzentrum f{\"u}r Schwerionenforschung, Mainz 55128, Germany}
\affiliation{Johannes Gutenberg-Universit{\"a}t Mainz, Mainz 55128, Germany}

\author{Deniz~Aybas\,\orcidlink{0000-0002-0392-5979}}
\affiliation{Department of Physics, Boston University, Boston, Massachusetts 02215, USA}
\affiliation{Department of Electrical and Computer Engineering, Boston University, Boston, Massachusetts 02215, USA}

\author{Hendrik~Bekker\,\orcidlink{0000-0002-6535-696X}}
\affiliation{Helmholtz-Institut, GSI Helmholtzzentrum f{\"u}r Schwerionenforschung, Mainz 55128, Germany}
\affiliation{Johannes Gutenberg-Universit{\"a}t Mainz, Mainz 55128, Germany}

\author{Dmitry~Budker\,\orcidlink{0000-0002-7356-4814}}
\affiliation{Helmholtz-Institut, GSI Helmholtzzentrum f{\"u}r Schwerionenforschung, Mainz 55128, Germany}
\affiliation{Johannes Gutenberg-Universit{\"a}t Mainz, Mainz 55128, Germany}
\affiliation{Department of Physics, University of California, Berkeley, California 94720, USA}

\author{Gary~P.~Centers\,\orcidlink{0000-0002-3798-0343}}
\affiliation{Helmholtz-Institut, GSI Helmholtzzentrum f{\"u}r Schwerionenforschung, Mainz 55128, Germany}
\affiliation{Johannes Gutenberg-Universit{\"a}t Mainz, Mainz 55128, Germany}

\author{Nataniel~L.~Figueroa\,\orcidlink{0000-0001-7703-1129}}
\affiliation{Helmholtz-Institut, GSI Helmholtzzentrum f{\"u}r Schwerionenforschung, Mainz 55128, Germany}
\affiliation{Johannes Gutenberg-Universit{\"a}t Mainz, Mainz 55128, Germany}

\author{Derek~F.~Jackson~Kimball\,\orcidlink{0000-0003-2479-6034}}
\affiliation{Department of Physics, California State University---East Bay, Hayward, California 94542, USA}

\author{Alexander~O.~Sushkov\,\orcidlink{0000-0001-8895-6338}}
\affiliation{Department of Physics, Boston University, Boston, Massachusetts 02215, USA}
\affiliation{Department of Electrical and Computer Engineering, Boston University, Boston, Massachusetts 02215, USA}
\affiliation{Photonics Center, Boston University, Boston, Massachusetts 02215, USA}

\begin{abstract}
We derive spectral line shapes of the expected signal for a haloscope experiment searching for axionlike dark matter. The knowledge of these line shapes is needed to optimize an experimental design and data analysis procedure. We extend the previously known results for the axion-photon and axion-gluon couplings to the case of gradient (axion-fermion) coupling. A unique feature of the gradient interaction is its dependence not only on magnitudes but also on directions of velocities of galactic halo particles, which leads to the directional sensitivity of the corresponding haloscope. We also discuss the daily and annual modulations of the gradient signal caused by the Earth's rotational and orbital motions. In the case of detection, these periodic modulations will be an important confirmation that the signal is sourced by axionlike particles in the halo of our Galaxy.
\end{abstract}

\maketitle

\section{Introduction}
\label{sec:intro}

According to diverse astronomical observations, about 85\% of the total mass of the Universe can be attributed to dark matter (DM), whose origin and composition remain unknown~\cite{Bertone_2005, Feng_2010, PDG_2020}. Most galaxies are thought to be embedded in DM halos, which play a key role in their formation and evolution~\cite{White_1978, Wechsler_2018}. Among the best-motivated DM candidates are the quantum chromodynamics axion and other light pseudoscalar bosons, which are collectively referred to as axionlike particles (ALPs)~\cite{Graham_2015, Irastorza_2018, Sikivie_2021}. Their characteristic feature is low mass ($m_a \ll 1~\text{eV}/c^2$) that leads to high number density. This feature distinguishes ALPs from other popular DM candidates, such as weakly interacting massive particles (WIMPs), which are much heavier. On the scale of laboratory detectors, ALPs exhibit wavelike, rather than particlelike, behavior. To first approximation, axionlike DM can be described as a classical field,
\begin{equation}
a(t) = a_0 \cos{(2\pi \nu_a t)}, \label{eq:toy_model}
\end{equation}
permeating space and oscillating at the ALP Compton frequency, $\nu_a = m_a c^2 / h$, where $c$ is the speed of light and $h = 2\pi \hbar$ is the Planck constant. The amplitude $a_0$ of the oscillations is related to the local DM energy density, $\rho_{\text{DM}}$, as $a_0 = \hbar \sqrt{2\rho_{\text{DM}}} / (m_a c)$~\cite{Graham_2013}. The canonical value of $\rho_{\text{DM}}$ is $0.3~\text{GeV}/\text{cm}^3$, which is accurate within a factor of 2--3~\cite{PDG_2020}.

Besides the gravitational interaction, there are three possible couplings between ALPs and Standard Model particles~\cite{Graham_2013}: (1) the axion-photon (or electromagnetic) coupling that mixes ALPs and photons, (2) the axion-gluon coupling giving rise to oscillating nuclear electric dipole moments, and (3)~the axion-fermion coupling between ALPs and nuclear or electron spins. The first two couplings, which are proportional to $a(t)$, are referred to as the ALP field couplings. The third one is proportional to the spatial gradient of $a(t)$ and is therefore referred to as the gradient coupling.

All three couplings listed above are used to search for ALPs in the DM halo of our Galaxy. The corresponding terrestrial detectors are usually called ``haloscopes'' to distinguish them from ``helioscopes'' looking for ALPs produced in the Sun~\cite{Sikivie_1983}. The axion-photon interaction is the most commonly targeted, but the other two couplings are also promising~\cite{Graham_2013}. Regardless of the chosen interaction, the knowledge of the expected signal line shape is needed to optimize any experimental design and data analysis procedure. Although the line shape for the axion-photon coupling has been known for decades~\cite{Krauss_1985, Turner_1990} and used for data analysis in multiple experiments (e.g., ADMX~\cite{Du_2018, Braine_2020}, CAPP~\cite{Lee_2020, Kwon_2021}, HAYSTAC~\cite{Brubaker_PRD_2017, Brubaker_PRL_2017, Backes_2021}, and SHAFT~\cite{Gramolin_2021}), there are no studies of its gradient counterpart. We fill this gap and derive spectral line shapes for both the ALP field and the gradient couplings using the same unified approach. We also discuss the daily and annual modulations of the gradient line shape, which, if detected, will be an important confirmation that the signal is sourced by ALPs in the halo of our Galaxy.

\section{Stochastic model of the ALP field}
\label{sec:stochastic_model}

Equation~\eqref{eq:toy_model} is only an approximate model for the field $a(t)$: it assumes that all ALPs in the galactic halo oscillate coherently and that the corresponding spectral line shape is a delta function $\delta(\nu - \nu_a)$. A more realistic model should account for the speed distribution of halo particles, which leads to a broadening of the line shape. This broadening occurs because frequencies of moving ALPs, as seen by an external observer, are larger than $\nu_a$ by an amount proportional to their kinetic energies:
\begin{equation}
\nu_n = \left(1 + \frac{v_n^2}{2c^2}\right) \nu_a, \label{eq:nu_n}
\end{equation}
where $\nu_n$ is the frequency of the $n$th particle and $v_n \ll c$ is its speed relative to the observer. Another effect spoiling coherence is that oscillations of different ALPs may not be synchronized. In this paper, we follow the most common assumption that their phases are completely uncorrelated.

The ALP field can be modeled more accurately as a superposition of $N$ independent oscillators~\cite{Foster_2018}:
\begin{equation}
a(\mathbf{r}, t) = \frac{a_0}{\sqrt{N}} \sum\limits_{n=1}^{N} \cos{(2\pi \nu_n t - \mathbf{k}_n \cdot \mathbf{r} + \phi_n)}, \label{eq:axion_field}
\end{equation}
where $\mathbf{k}_n = m_a \mathbf{v}_n / \hbar$ is the wave vector of the $n$th ALP, $\mathbf{v}_n$ is its velocity, and the phases $\phi_n \in [0, \, 2\pi)$ are uniformly distributed. The velocities~$\mathbf{v}_n$ are sampled from the velocity distribution of halo particles. The frequencies~$\nu_n$ are given by Eq.~\eqref{eq:nu_n} with $v_n = |\mathbf{v}_n|$. The model~\eqref{eq:axion_field} is similar to that describing chaotic light with Doppler broadening~\cite{Loudon}.

It is instructive to qualitatively discuss the effects caused by different terms in the cosine argument in Eq.~\eqref{eq:axion_field}. As already mentioned, the first term, $2\pi \nu_n t$, leads to the broadening of the spectral line shape. To experimentally resolve the line shape, one needs to have sufficiently long interrogation time~$T$ compared to the ALP coherence time~$\tau_c$. Therefore, this effect is important when $T \gg \tau_c$, which can be rewritten in a form useful for quick estimates as $T[\text{s}] \gg 2 / m_a[\text{neV}/c^2]$. Note that the sensitivity to the ALP coupling scales with $T$ as $T^{1/2}$ when $T \ll \tau_c$ (coherent averaging) and as $(\tau_c T)^{1/4}$ when $T \gg \tau_c$ (incoherent averaging)~\cite{Budker_2014}. While the majority of haloscope searches for the axion-photon coupling~\cite{Du_2018, Braine_2020, Lee_2020, Kwon_2021, Brubaker_PRD_2017, Brubaker_PRL_2017, Backes_2021, Gramolin_2021} have operated in the regime where $T \gg \tau_c$, many of the experiments targeting the gradient coupling~\cite{Abel_2017, Terrano_2019, Garcon_2019, Wu_2019, Jiang_2021, Bloch_2021} have operated in the $T < \tau_c$ regime, more amenable to a time-domain analysis such as that presented in Refs.~\cite{Centers_2019, Lisanti_2021}. However, the CASPEr~\cite{Aybas_2021} and QUAX~\cite{Crescini_2018, Crescini_2020} experiments, for example, are now exploring the $T \gg \tau_c$ regime, where knowledge of the line shape as discussed here can be important for data analysis.

The term $\mathbf{k}_n \cdot \mathbf{r}$ can be eliminated, in the case of a single detector sensitive to the ALP field couplings, by choosing the coordinate system with $\mathbf{r} = 0$. In contrast, two detectors located at positions $\mathbf{r}_1$ and $\mathbf{r}_2$ lead to the nonvanishing term $\mathbf{k}_n \cdot (\mathbf{r}_1 - \mathbf{r}_2)$. Therefore, an experiment exploring correlations between two or more spatially separated detectors can probe the three-dimensional velocity distribution of halo ALPs rather than the speed distribution~\cite{Derevianko_2018, Foster_2021}. The same result can be achieved with only a single detector sensitive to the gradient coupling. This is because $\nabla a$ sourced by each ALP is proportional to its velocity $\mathbf{v}_n$, as can be seen after calculating the gradient of the field~\eqref{eq:axion_field}:
\begin{equation}
\nabla a(\mathbf{r}, t) = \frac{\sqrt{2\rho_{\text{DM}}}}{c \sqrt{N}} \sum\limits_{n=1}^{N} \mathbf{v}_n \sin{(2\pi \nu_n t - \mathbf{k}_n \cdot \mathbf{r} + \phi_n)}. \label{eq:nabla_a}
\end{equation}
Note that the model~\eqref{eq:axion_field} assumes that the amplitude $a_0$ does not have any spatial dependence, which corresponds to a homogeneous ALP field. In the most general case, spatial inhomogeneity of the field $a(\mathbf{r}, t)$ also contributes to the gradient~\eqref{eq:nabla_a}.

Having different arguments $(2\pi \nu_n t + \phi_n)$, the cosine waves in Eq.~\eqref{eq:axion_field} interfere with each other, which manifests in stochastic fluctuations of the ALP field amplitude~\cite{Centers_2019}. This effect is similar to acoustic beats caused by interference between multiple tones of slightly different frequencies. The resulting stochastic amplitudes follow a Rayleigh distribution, as discussed in Sec.~\ref{sec:statistics}, Appendix~\ref{sec:Appendix_A}, and Refs.~\cite{Foster_2018, Centers_2019}. In the next two sections, we leave these stochastic fluctuations aside and derive statistically averaged line shapes for both types of couplings.

\section{The case of ALP field couplings}

In this section, we consider the case of axion-photon or axion-gluon couplings and show how the corresponding spectral line shape can be derived. For these couplings, the detector response (e.g., the voltage induced in a pickup coil) is proportional to either the ALP field $a(t)$ itself or its time derivative. The experimentalist records this response, $s(t)$, for a long interrogation time, $T \gg \tau_c$. The raw time-domain data are then converted to the frequency domain by calculating their Fourier transform, $S(\nu)$. The most convenient quantity to analyze is the power spectral density (PSD, also called power spectrum), which is $|S(\nu)|^2$. The PSD shows how the average power of the signal is distributed over the frequency~$\nu$. It satisfies Parseval's theorem
\begin{equation}
P = \frac{1}{T} \int\limits_{0}^{T} |s(t)|^2 \, dt = \int\limits_{0}^{\infty} |S(\nu)|^2 \, d\nu, \label{eq:Parseval}
\end{equation}
where $P$ is the signal power averaged over the interrogation time~$T$. The spectral line shape, $\lambda(\nu)$, is a closely related quantity, defined as $\lambda(\nu) = |S(\nu)|^2/P$, so that it is normalized to unity:
\begin{equation}
\int\limits_{0}^{\infty} \lambda(\nu) \, d\nu = 1.
\label{eq:norm}
\end{equation}

To derive the line shape, we assume a continuous limit, $N \rightarrow \infty$, of the discrete model for the ALP field discussed in Sec.~\ref{sec:stochastic_model}. Then, Eq.~\eqref{eq:nu_n} can be rewritten as
\begin{equation}
v(\nu) = c\sqrt{2(\nu / \nu_a - 1)}. \label{eq:v_nu}
\end{equation}
Equation~\eqref{eq:v_nu} suggests that $\lambda(\nu)$ can be obtained from the distribution function, $f(v)$, of ALP speeds in the halo by changing variables from~$v$ to~$\nu$:
\begin{equation}
\left. \lambda(\nu) = f(v) \, \frac{dv}{d\nu} \right|_{v = c \sqrt{2(\nu / \nu_a - 1)}}. \label{eq:lambda_nu}
\end{equation}
In the case of gradient coupling, Eq.~\eqref{eq:lambda_nu} involves a more complicated distribution function that accounts for the spatial orientation of the detector (see Sec.~\ref{sec:gradient}).

As is typical for direct-detection experiments, we assume the standard halo model for the DM halo of our Galaxy~\cite{Schumann_2019, Evans_2019}. According to this model, the velocities $\mathbf{v}$ of DM particles in the galactic rest frame follow the Maxwell-Boltzmann distribution
\begin{equation}
f_{\text{gal}}^{(3)}(\mathbf{v}) \, d^3 \mathbf{v} = \frac{1}{\pi^{3/2} v_0^3} \exp{\left(-\frac{\mathbf{v}^2}{v_0^2}\right)} \, d^3 \mathbf{v}, \label{eq:f_gal}
\end{equation}
where $v_0 \approx 220~\text{km}/\text{s}$ is the circular rotation speed of the Galaxy at the solar radius~\cite{Evans_2019}. To clearly distinguish between one-dimensional and three-dimensional distribution functions, we denote them as $f(v)$ and $f^{(3)}(\mathbf{v})$, respectively. Note that DM particles moving faster than the escape speed, $v_{\text{esc}} \approx 544~\text{km}/\text{s}$, are not bound by the gravitational potential of the Galaxy. Therefore, the distribution~\eqref{eq:f_gal} should be truncated at $|\mathbf{v}| > v_{\text{esc}}$, but this effect leads to only minor corrections that are not significant for our analysis.

The velocity distribution~\eqref{eq:f_gal} should be modified to account for the fact that any Earth-based laboratory moves through the DM halo with a relative velocity~$\mathbf{v}_{\text{lab}}$:
\begin{equation}
f_{\text{lab}}^{(3)}(\mathbf{v}) = f_{\text{gal}}^{(3)}(\mathbf{v} - \mathbf{v}_{\text{lab}}). \label{eq:f_lab_velocity}
\end{equation}
The velocity~$\mathbf{v}_{\text{lab}}$ is dominated by the Sun's motion relative to the galactic frame at the speed $v_{\odot} \approx 233~\text{km}/\text{s}$. However, both the magnitude and the direction of~$\mathbf{v}_{\text{lab}}$ are periodically modulated due to the orbital and rotational motions of the Earth. These modulations are considered in Sec.~\ref{sec:modulations}, but for now we assume that $\mathbf{v}_{\text{lab}}$ is fixed.

\begin{figure}
\includegraphics[width=0.6\columnwidth]{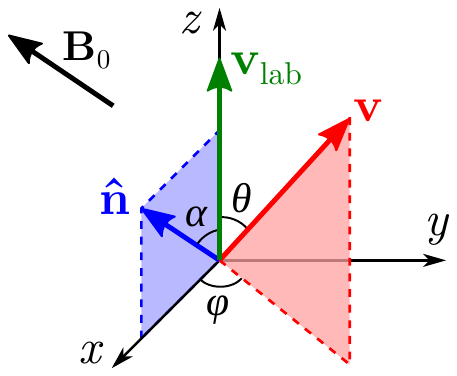}
\caption{\label{fig:coordinates} The coordinate system used in this paper. The following vectors are shown: $\mathbf{v}_{\text{lab}} = (0, \, 0, \, v_{\text{lab}})$ is the laboratory velocity relative to the galactic rest frame, $\mathbf{v} = (v \sin{\theta} \cos{\phi}, \, v \sin{\theta} \sin{\phi}, \, v \cos{\theta})$ is the velocity of an individual ALP in the galactic rest frame, and $\mathbf{\hat n} = (\sin{\alpha}, \, 0, \, \cos{\alpha})$ is the unit vector directed along the external static magnetic field~$\mathbf{B}_0$.}
\end{figure}

To derive the lab-frame speed distribution, $f_{\text{lab}}(v)$, from the velocity distribution~\eqref{eq:f_lab_velocity}, we employ spherical coordinates $(v, \, \theta, \, \phi)$ chosen such that $v = |\mathbf{v}|$, the $z$~axis is directed along~$\mathbf{v}_{\text{lab}}$, and the polar angle~$\theta$ is the angle between $\mathbf{v}$ and~$\mathbf{v}_{\text{lab}}$ (see Fig.~\ref{fig:coordinates}). Then, taking into account that $d^3 \mathbf{v} = v^2 \, dv \, \sin{\theta} \, d\theta \, d\phi$, we can write $f_{\text{lab}}(v)$ as the following integral over the angles $\theta$ and $\phi$:
\begin{equation}
f_{\text{lab}}(v) \, dv = v^2 \, dv \int\limits_{0}^{2\pi} d\phi \int\limits_{0}^{\pi} f_{\text{lab}}^{(3)}(\mathbf{v}) \sin{\theta} \, d\theta. \label{eq:f_lab_integral}
\end{equation}
After substituting into Eq.~\eqref{eq:lambda_nu} the result of the integration~\eqref{eq:f_lab_integral} and the derivative of Eq.~\eqref{eq:v_nu}, we finally obtain the following spectral line shape:
\begin{gather}
\lambda(\nu) = \frac{2c^2}{\sqrt{\pi} v_0 v_{\text{lab}} \nu_a} \exp{\left(-\frac{\beta^2 v_0^2}{4v_{\text{lab}}^2} - \frac{v_{\text{lab}}^2}{v_0^2}\right)} \sinh{\beta}, \label{eq:non-gradient_lineshape}
\end{gather}
where $v_{\text{lab}} = |\mathbf{v}_{\text{lab}}|$ and we have denoted for brevity
\begin{equation}
\beta =\frac{2c v_{\text{lab}}}{v_0^2} \sqrt{\frac{2(\nu - \nu_a)}{\nu_a}}.
\end{equation}
In different forms, the line shape~\eqref{eq:non-gradient_lineshape} has been previously reported in Refs.~\cite{Turner_1990, O'Hare_2017, Derevianko_2018, Foster_2018}.

We can also obtain an expression for the corresponding power spectrum. Let us assume that the signal is $s(t) = \kappa a(t)$, where $a(t)$ is the ALP field and $\kappa$ is a factor proportional to the coupling strength and dependent on specific experimental details. Then, according to Parseval's theorem~\eqref{eq:Parseval},
\begin{equation}
|S(\nu)|^2 = \frac{1}{2} (\kappa a_0)^2 \lambda(\nu). \label{eq:S_non-gradient}
\end{equation}

The dimensionless line shape $\nu_a \lambda(\nu)$ is shown as a black dotted curve in Fig.~\ref{fig:lineshapes}(a). Its shape is highly asymmetric: there is a steep rise starting at the frequency $\nu_a$ and a long tail at high frequencies. We can also see that the spectral linewidth, defined as the full width at half maximum, is $\Delta \nu / \nu_a \approx v_0^2 / c^2 \approx 10^{-6}$. Assuming that the relationship between the linewidth $\Delta \nu$ and the coherence time $\tau_c$ is the same as for a Lorentzian~\cite{Loudon}, we can estimate
\begin{equation}
\tau_c = \frac{1}{\pi \Delta \nu} \approx \frac{2 \hbar}{m_a v_0^2}.
\end{equation}
Note that there is an ambiguity, up to a factor of $2\pi$, in the definitions of $\tau_c$ used in the literature~\cite{Centers_2019}.

\section{The case of gradient coupling}
\label{sec:gradient}

The axion-fermion (or gradient) coupling to nuclear spins can be described by the nonrelativistic Hamiltonian
\begin{equation}
H = \hbar c g \, \nabla a \cdot \mathbf{I}, \label{eg:Hamiltonian_gr}
\end{equation}
where $g$ is the coupling strength, $\nabla a$ is the spatial gradient of the ALP field, and $\mathbf{I}$ is the nuclear spin operator~\cite{Graham_2013}. Note that the factor $\hbar c$ in Eq.~(\ref{eg:Hamiltonian_gr}) is written assuming that the combination $g a_0$ is dimensionless. By drawing an analogy with the Zeeman effect, we can think of $\nabla a$ as a pseudomagnetic field oscillating at the frequency~$\nu_a$. There are different experimental approaches for detecting this field. For example, the CASPEr-ZULF experiments~\cite{Garcon_2019, Wu_2019} search for ALP-induced modulations of Zeeman splittings between nuclear energy levels in an ultralow external magnetic field~$\mathbf{B}_0$. Since these experiments measure small perturbations of the leading magnetic field, they are sensitive only to the component of $\nabla a$ parallel to~$\mathbf{B}_0$. Therefore, the time-domain signal can be written in this case as $s_{\parallel}(t) = \kappa_{\parallel} \nabla a_{\parallel}(t)$, where the factor $\kappa_{\parallel}$ depends on the coupling strength and specific experimental details.

The CASPEr-Gradient and CASPEr-Electric experiments use a different approach~\cite{Budker_2014, Aybas_2021}. An ensemble of nuclear spins is initially polarized in a strong magnetic field~$\mathbf{B}_0$ so that the net magnetization of the sample is parallel to~$\mathbf{B}_0$. The ALP-induced pseudomagnetic field serves as an oscillating driving field. If the Larmor frequency of the nuclear spins [$\nu_L = \gamma B_0 / (2\pi)$, where $\gamma$ is the gyromagnetic ratio of the nuclei] matches the frequency~$\nu_a$ of the driving field, then a magnetic resonance occurs, resulting in a torque on the nuclear spins. This torque causes the spins to precess around the $\mathbf{B}_0$ axis, which leads to an oscillating transverse magnetization of the sample. One can use a sensitive magnetometer to detect this transverse magnetization. This experimental technique is sensitive only to the component of $\nabla a$ perpendicular to~$\mathbf{B}_0$. Therefore, the signal in this case is $s_{\perp}(t) = \kappa_{\perp} \nabla a_{\perp}(t)$. Note that CASPEr-Electric~\cite{Aybas_2021} is sensitive to both the axion-gluon and the axion-fermion couplings, but here we focus on the latter, since the signal due to the axion-gluon coupling has the line shape~\eqref{eq:non-gradient_lineshape}.

In the rest of this section, we derive spectral line shapes and power spectra for the two types of experiments described above. We characterize the direction of the leading magnetic field~$\mathbf{B}_0$ by the unit vector $\mathbf{\hat n} = \mathbf{B}_0 / B_0$ (see Fig.~\ref{fig:coordinates}). As follows from Eq.~\eqref{eq:nabla_a}, each ALP produces a $\nabla a$ that is proportional to its velocity~$\mathbf{v}$. Therefore, to take into account the directional sensitivity of the detector, we first determine the components of $\mathbf{v}$ parallel and perpendicular to the leading field:
\begin{equation}
v_{\parallel} = \mathbf{v} \cdot \mathbf{\hat n}, \qquad v_{\perp} = \sqrt{v^2 - v_{\parallel}^2}. \label{eq:v_components}
\end{equation}
Then, we integrate over the angles $\theta$ and $\phi$ in the same way as in Eq.~\eqref{eq:f_lab_integral} but with the integrand multiplied by each of the squared components~\eqref{eq:v_components}:
\begin{equation}
f_{\parallel, \perp}(v) \, dv = \frac{v^2 \, dv}{C_{\parallel, \perp}} \int\limits_{0}^{2\pi} d\phi \int\limits_{0}^{\pi} v_{\parallel, \perp}^2 f_{\text{lab}}^{(3)}(\mathbf{v}) \sin{\theta} \, d\theta, \label{eq:f_lab_integration}
\end{equation}
where the normalization coefficients,
\begin{equation}
C_{\parallel} = \frac{v_0^2}{2} + v_{\text{lab}}^2 \cos^2{\alpha}, \quad C_{\perp} = v_0^2 + v_{\text{lab}}^2 \sin^2{\alpha}, \label{eq:norm_coefficients}
\end{equation}
which depend on the angle~$\alpha$ between the vectors $\mathbf{\hat n}$ and $\mathbf{v}_{\text{lab}}$, are chosen so that
\begin{equation}
\int\limits_{0}^{\infty} f_{\parallel, \perp}(v) \, dv = 1.
\end{equation}
The factors $v_{\parallel}^2$ and $v_{\perp}^2$ appear in Eq.~\eqref{eq:f_lab_integration} because the corresponding PSDs are proportional to $|\nabla a_{\parallel}|^2$ and $|\nabla a_{\perp}|^2$, respectively. The normalization coefficients~\eqref{eq:norm_coefficients} ensure that $f_{\parallel}(v)$ and $f_{\perp}(v)$ are proper distribution functions.

After calculating $f_{\parallel}(v)$ and substituting the result into Eq.~\eqref{eq:lambda_nu}, we obtain the following spectral line shape for experiments sensitive to the parallel component of the gradient:
\begin{gather}
\lambda_{\parallel}(\nu) = \lambda(\nu) \, \frac{2c^2}{C_{\parallel}} \frac{\nu - \nu_a}{\nu_a} \nonumber \\
{} \times \left[\cos^2{\alpha} - \frac{1}{\beta} \left(\coth{\beta} - \frac{1}{\beta}\right) \left(2 - 3\sin^2{\alpha}\right)\right]. \label{eq:lambda_par}
\end{gather}
Similarly, by repeating the same calculation for $f_{\perp}(v)$, we derive the line shape for the case of magnetic resonance experiments sensitive to $\nabla a_{\perp}$:
\begin{gather}
\lambda_{\perp}(\nu) = \lambda(\nu) \, \frac{2c^2}{C_{\perp}} \frac{\nu - \nu_a}{\nu_a} \nonumber \\
{} \times \left[\sin^2{\alpha} + \frac{1}{\beta} \left(\coth{\beta} - \frac{1}{\beta}\right) \left(2 - 3\sin^2{\alpha}\right)\right]. \label{eq:lambda_perp}
\end{gather}

\begin{figure*}
\includegraphics[width=\textwidth]{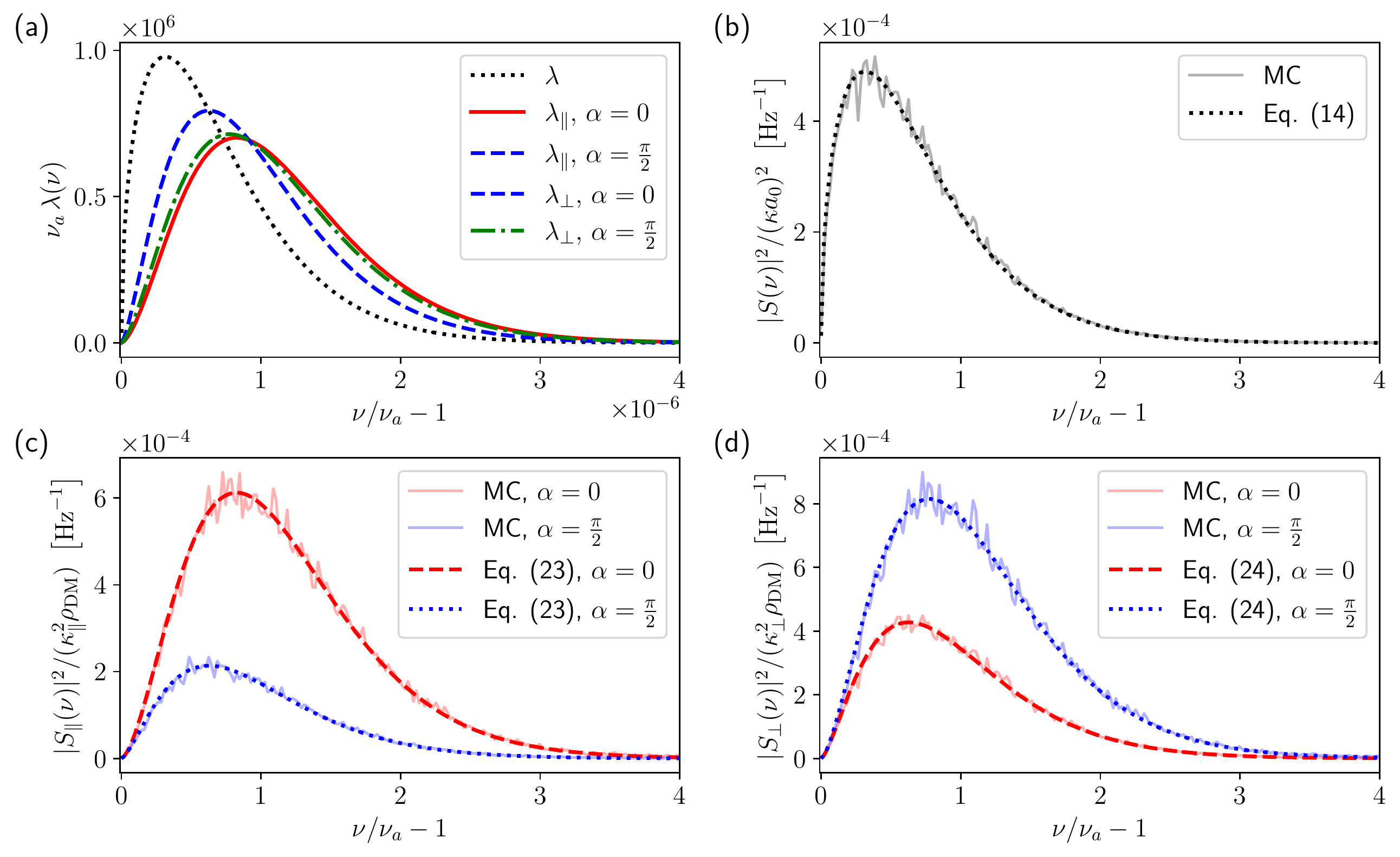}
\caption{\label{fig:lineshapes}Spectral line shapes and power spectra for both types of ALP couplings. (a)~Dimensionless line shapes $\nu_a \lambda(\nu)$, where $\lambda(\nu)$ are given by Eqs.~\eqref{eq:non-gradient_lineshape}, \eqref{eq:lambda_par}, and \eqref{eq:lambda_perp}. For the gradient coupling, we consider two detector orientations: $\alpha = 0$ and $\alpha = \pi/2$. Note that the line shape $\lambda_{\perp},~\alpha = 0$ coincides with $\lambda_{\parallel},~\alpha = \pi/2$. (b)~Monte Carlo simulation (solid curve) and analytical description (dotted curve) for the power spectrum in the case of ALP field couplings. Monte Carlo results are obtained by averaging over 500 PSDs calculated using a Fourier transform of the time-domain signal $s(t) = \kappa a(t)$, where $a(t)$ is given by Eq.~\eqref{eq:axion_field}. Note that we increased the linewidth by a factor of $10^6$ to reduce the computational cost of the simulation. The analytical PSD is given by Eq.~\eqref{eq:S_non-gradient}. (c)~Similar to panel~(b) but for the parallel gradient case. Two detector orientations are considered: $\alpha = 0$ and $\alpha = \pi/2$. The time-domain signal is $s(t) = \kappa_{\parallel} \nabla a_{\parallel}(t)$, where $\nabla a(t)$ is given by Eq.~\eqref{eq:nabla_a}. The analytical PSD is given by Eq.~\eqref{eq:S_par}. (d) Similar to panels (b) and (c) but for the perpendicular gradient case. The time-domain signal is $s(t) = \kappa_{\perp} \nabla a_{\perp}(t)$, and the analytical PSD is given by Eq.~\eqref{eq:S_perp}.}
\end{figure*}

The dimensionless quantities $\nu_a \lambda_{\parallel}(\nu)$ and $\nu_a \lambda_{\perp}(\nu)$ are shown in Fig.~\ref{fig:lineshapes}(a) for two spatial orientations of the detector ($\alpha = 0$ and $\alpha = \pi / 2$). We can see that the gradient line shapes rise slower and reach maxima at higher frequencies than the curve~\eqref{eq:non-gradient_lineshape}. There is also a noticeable dependence of the gradient line shapes on the angle~$\alpha$.

After taking into account Parseval's theorem~\eqref{eq:Parseval}, we obtain the following expressions for the corresponding PSDs:
\begin{align}
|S_{\parallel}(\nu)|^2 &= P_{\parallel} \lambda_{\parallel}(\nu), \label{eq:S_par} \\ 
|S_{\perp}(\nu)|^2 &= P_{\perp} \lambda_{\perp}(\nu), \label{eq:S_perp}
\end{align}
where
\begin{equation}
P_{\parallel, \perp} = \frac{\rho_{\text{DM}}}{c^2} \kappa_{\parallel, \perp}^2 C_{\parallel, \perp}
\end{equation}
is the total signal power. Note that, as can be seen from Eq.~\eqref{eq:v_components}, $P_{\parallel}$ results from a projection along the axis defined by $\mathbf{\hat{n}}$, while $P_{\perp}$ results from a projection into the plane orthogonal to $\mathbf{\hat{n}}$. In the parallel case, the signal power is maximum when $\alpha = 0$ and minimum when $\alpha = \pi / 2$. In the perpendicular case, it is vice versa. The ratio of the maximum to the minimum values is $\approx 3$ for $P_{\parallel}$ and $\approx 2$ for $P_{\perp}$ (assuming that $v_0 \approx v_{\text{lab}}$).

\section{Statistics of the ALP field}
\label{sec:statistics}

The line shapes and power spectra derived in the previous sections are smooth functions of frequency $\nu$. However, the ALP field~\eqref{eq:axion_field} is a stochastic variable in the sense that its amplitude and phase vary randomly on a timescale of~$\tau_c$. As a consequence, even for a noiseless experiment sensitive to axionlike DM, the power spectrum is a stochastic function of $\nu$~\cite{Foster_2018, Centers_2019, Lisanti_2021}. The expressions \eqref{eq:S_non-gradient}, \eqref{eq:S_par}, and \eqref{eq:S_perp} are the expected values for the corresponding PSDs. In this section, we discuss the statistical properties of both the ALP field and the resulting power spectra.\footnote{While we have not carried out a detailed comparison of results, our treatment here is based on the same assumptions as the approaches described in Refs.~\cite{Foster_2018, Centers_2019, Lisanti_2021} to analyze the stochastic properties of ALP signals. In particular, our power spectra should correspond, in the regime $T \gg \tau_c$, to Fourier transforms of the correlation functions obtained in Ref.~\cite{Lisanti_2021}. Although the authors of Ref.~\cite{Lisanti_2021} point out some differences in their approach with respect to that described in a preprint of Ref.~\cite{Centers_2019}, it turns out that these differences were corrected for in the published version of Ref.~\cite{Centers_2019}.}

Before proceeding, let us make a digression into probability theory to introduce results important for the subsequent discussion (see Appendix~\ref{sec:Appendix_A} for further details). Consider a random variable~$z$ defined as the sum,
\begin{equation}
z = \sum\limits_{n=1}^{N} s_n \exp{(i \phi_n)}, \label{eq:random_walk}
\end{equation}
of complex numbers with random magnitudes~$s_n$ and arguments~$\phi_n$. The $s_n$ values are drawn from some probability distribution with mean $\mu_s$ and variance $\sigma_s^2$, while the phases $\phi_n \in [0, \, 2\pi)$ follow a uniform distribution. The summation~\eqref{eq:random_walk} corresponds to a two-dimensional random walk on the complex plane: we start at the origin and make $N$ steps, each of size~$s_n$ and in the direction given by the angle~$\phi_n$. Then, the complex number $z = x + i y$ specifies the $(x, \, y)$ coordinates of the end point. As shown in Appendix~\ref{sec:Appendix_A} using the central limit theorem, both $x$ and $y$ are drawn from the normal distribution with zero mean and variance $\sigma^2 = N (\mu_s^2 + \sigma_s^2) / 2$. This result does not depend on the specific probability distribution for $s_n$ as long as $N \gg 1$.

It is also instructive to rewrite $z$ in polar form as $z = r \exp{(i \phi')}$. Then, the absolute value $r = \sqrt{x^2 + y^2}$ represents the distance from the origin and follows the Rayleigh distribution with probability density function
\begin{equation}
p \, (r; \, \sigma) = \frac{r}{\sigma^2} \exp{\left(-\frac{r^2}{2\sigma^2}\right)}. \label{eq:Rayleigh_distribution}
\end{equation}
The argument $\phi'$ follows a uniform distribution, which reflects the isotropy of the random walk.

The above results allow us to evaluate the sum of cosine waves having the same frequency~$\nu$ but random amplitudes~$s_n$ and phases~$\phi_n$ as
\begin{gather}
\sum\limits_{n=1}^{N} s_n \cos{(2\pi \nu t + \phi_n)} \nonumber \\
= \sqrt{\frac{N (\mu_s^2 + \sigma_s^2)}{2}} \, r' \cos{(2\pi \nu t + \phi')}, \label{eq:summation_cos}
\end{gather}
where $\phi' \in [0, \, 2\pi)$ is drawn from a uniform distribution and $r'$ is drawn from the Rayleigh distribution~\eqref{eq:Rayleigh_distribution} with $\sigma = 1$. Equation~\eqref{eq:summation_cos} is derived in Appendix~\ref{sec:Appendix_A}, and a similar relation also holds for sine waves. Note that the variable $r' = r / \sigma$ in Eq.~\eqref{eq:summation_cos} is a dimensionless version of the distance~$r$ discussed in the previous paragraph.

We are now ready to consider the case of ALP field couplings and calculate the sum~\eqref{eq:axion_field} over all $N$ terms. To make this calculation feasible, we partition the full set of $N$ particles into subsets labeled by index~$j$ and containing $N_j$ ALPs with lab-frame speeds between $v_j$ and $v_j + \Delta v$, where $\Delta v$ is a small interval~\cite{Foster_2018}. The contribution of the $j$th subset to the ALP field $a(t)$ can be evaluated as
\begin{gather}
a_j(t) = \frac{a_0}{\sqrt{N}} \sum\limits_{n=1}^{N_j} \cos{(2\pi \nu_j t + \phi_n)} \nonumber \\
= \frac{a_0}{\sqrt{N}} \sqrt{\frac{N_j}{2}} \, r'_j \cos{\left(2\pi \nu_j t + \phi'_j\right)},
\end{gather}
where the first line follows from Eq.~\eqref{eq:axion_field} with $\mathbf{r} = 0$ and the summation is performed using Eq.~\eqref{eq:summation_cos} with $s_n = 1$, $\mu_s = 1$, and $\sigma_s = 0$, which corresponds to a random walk with unit step size. After summing over all the subsets, we finally obtain
\begin{equation}
a(t) = \frac{a_0}{\sqrt{2}} \sum\limits_j \sqrt{f_{\text{lab}}(v_j) \Delta v} \, r'_j \cos{\left(2\pi \nu_j t + \phi'_j\right)}, \label{eq:a_stochastic}
\end{equation}
where we have taken into account that the number of ALPs in the $j$th subset is $N_j = N f_{\text{lab}}(v_j) \Delta v$ with $f_{\text{lab}}(v)$ given by Eq.~\eqref{eq:f_lab_integral}.

Each term in the sum~\eqref{eq:a_stochastic} corresponds to the subset containing ALPs with speeds $\approx v_j$ and frequencies $\approx \nu_j$, where $\nu_j$ is given by Eq.~\eqref{eq:nu_n}. The relative contribution of each subset is governed by two factors: $\sqrt{f_{\text{lab}}(v_j)}$ and $r'_j$. The former, deterministic factor describes the expected value of the field amplitude at the frequency~$\nu_j$ and is related to the spectral line shape $\lambda(\nu_j)$ by Eq.~\eqref{eq:lambda_nu}. The latter, stochastic factor is drawn from the Rayleigh distribution~\eqref{eq:Rayleigh_distribution} with $\sigma = 1$. It is in this sense that the amplitude of the ALP field is a Rayleigh-distributed stochastic variable.

Since the power contained in the $j$th subset is proportional to $|a_j(t)|^2$, it is also a stochastic variable distributed as $(r'_j)^2$. As discussed in Appendix~\ref{sec:Appendix_A}, the square of a Rayleigh-distributed variable follows an exponential distribution. Therefore, the PSD at each frequency~$\nu_j$ is drawn from the exponential distribution with probability density function
\begin{equation}
p\left(|S_j(\nu_j)|^2\right) = \frac{1}{|S(\nu_j)|^2} \exp{\left(-\frac{|S_j(\nu_j)|^2}{|S(\nu_j)|^2}\right)}, \label{eq:exponential_distribution}
\end{equation}
where $|S(\nu_j)|^2$ is the expected value given by Eq.~\eqref{eq:S_non-gradient}.

The same approach can be extended to the case of gradient coupling. We skip the intermediate steps and provide here only the final result for the parallel and perpendicular components of the ALP field gradient:
\begin{gather}
\nabla a_{\parallel, \perp} (t) = \frac{\sqrt{\rho_{\text{DM}}}}{c} \sum\limits_j \left[\sqrt{\mu_{\parallel, \perp}^2(v_j) + \sigma_{\parallel, \perp}^2(v_j)} \rule{0mm}{5mm}\right. \nonumber \\
\times \left. \sqrt{f_{\text{lab}}(v_j) \Delta v} \, r'_j \sin{\left(2\pi \nu_j t + \phi'_j\right)}\right], \label{eq:nabla_a_stochastic}
\end{gather}
where $\mu_{\parallel, \perp}(v_j)$ and $\sigma_{\parallel, \perp}^2(v_j)$ are the mean and variance of $v_{\parallel, \perp}$ for the $j$th subset, see Eqs.~\eqref{eq:mean} and~\eqref{eq:variance} for explicit definitions. As shown in Appendix~\ref{sec:Appendix_B},
\begin{equation}
\bigl(\mu_{\parallel, \perp}^2 + \sigma_{\parallel, \perp}^2\bigr) f_{\text{lab}}(v_j) = C_{\parallel, \perp} f_{\parallel, \perp}(v_j), \label{eq:deterministic_factor}
\end{equation}
with $f_{\parallel, \perp}(v)$ given by Eq.~\eqref{eq:f_lab_integration}. Since the factor~\eqref{eq:deterministic_factor} is deterministic and the factor $r'_j$ in Eq.~\eqref{eq:nabla_a_stochastic} is again Rayleigh distributed, the gradient coupling has the same statistical properties as the nongradient couplings considered above. In particular, the power spectra follow the exponential distribution~\eqref{eq:exponential_distribution}, with the expected values given by Eqs.~\eqref{eq:S_par} and~\eqref{eq:S_perp}.

To illustrate the stochastic nature of the ALP field and to verify our derivation of the power spectra, we performed a dedicated Monte Carlo simulation. In the nongradient case, we generated the time-domain signal $s(t) = \kappa a(t)$ using the model~\eqref{eq:axion_field} for $a(t)$ with $N = 10^3$ particles. We sampled ALP velocities according to the Maxwell-Boltzmann distribution~\eqref{eq:f_gal} and then used Eq.~\eqref{eq:f_lab_velocity} to transform them from the galactic rest frame to the laboratory frame. We assumed the following parameters: Compton frequency $\nu_a = 1~\text{kHz}$, sampling frequency of 10~kHz, and interrogation time $T = 0.05~\text{s}$. To reduce the computational cost of the simulation, we set $v_0 = 2.2 \times 10^5~\text{km}/\text{s}$ and $v_{\text{lab}} = 2.33 \times 10^{5}~\text{km}/\text{s}$, which increased the width of the spectral line by a factor of $10^6$ while preserving its characteristic shape. We calculated the power spectrum by performing a Fourier transform of the signal $s(t)$ and normalizing the result according to Parseval's theorem~\eqref{eq:Parseval}. We repeated this process 500 times and averaged over the individual PSDs in order to reduce the size of stochastic fluctuations. The resulting averaged power spectrum is shown in Fig.~\ref{fig:lineshapes}(b) in comparison with the analytical formula~\eqref{eq:S_non-gradient}.

For the gradient coupling, we followed the same procedure but assumed that $s(t) = \kappa_{\parallel} \nabla a_{\parallel}(t)$ in the parallel case and $s(t) = \kappa_{\perp} \nabla a_{\perp}(t)$ in the perpendicular case, where $\nabla a(t)$ is given by Eq.~\eqref{eq:nabla_a}. We obtained the corresponding projections of $\nabla a$ by substituting the velocity components~\eqref{eq:v_components} into Eq.~\eqref{eq:nabla_a} instead of~$\mathbf{v}$. The resulting averaged power spectra are shown in Figs.~\ref{fig:lineshapes}(c) and \ref{fig:lineshapes}(d) for the parallel and perpendicular cases, respectively, and for two spatial orientations of the detector ($\alpha = 0$ and $\alpha = \pi / 2$). For both types of couplings, there is good agreement between the Monte Carlo simulation and our analytical expressions. We also verified that the distribution of PSD values within each frequency bin matches the exponential distribution~\eqref{eq:exponential_distribution}. Despite the averaging, the simulated power spectra are stochastic and scattered around the expected values, as can be seen in Fig.~\ref{fig:lineshapes}. For further details on the simulation, we refer the reader to our \textsc{Python} code~\cite{GitHub}.

Finally, we note that there is an important difference between the gradient and the nongradient cases. As shown in Ref.~\cite{Centers_2019}, stochastic amplitude fluctuations can reduce the sensitivity of an axion-photon or an axion-gluon haloscope by as much as an order of magnitude in the regime with $T \ll \tau_c$. A gradient haloscope with three mutually orthogonal sensitivity axes is significantly less susceptible to this effect. This is because both the stochastic amplitude $r'_j$ and the phase $\phi'_j$ are independent random variables for each of the three axes and for each frequency $\nu_j$. Although we are unable to resolve different frequencies when $T \ll \tau_c$, we still have three independently sampled amplitudes. The probability of all three values being small is suppressed compared to the nongradient case of a single amplitude.

\section{Periodic modulations of the gradient signal}
\label{sec:modulations}

As already mentioned, the observer's velocity relative to the galactic DM halo is periodically modulated due to the Earth's orbital and rotational motions. It is well known that annual modulations in the event detection rate are an important experimental signature for WIMP searches~\cite{Drukier_1986, Freese_2013, Schumann_2019}. The same is true for axionlike DM experiments. One needs to know the time dependence of the power spectrum in order to optimize the detector sensitivity and data analysis procedure. In the case of detection, the periodic modulations will be an important confirmation that the signal is indeed sourced by the galactic DM halo. In this section, we consider annual and daily modulations of the axionlike DM signal in the case of gradient coupling.

Both the magnitude and the direction of the vector $\mathbf{v}_{\text{lab}}$ vary with time. There are several reasons why it is convenient to consider these time dependences separately. First, the line shape~\eqref{eq:non-gradient_lineshape} is sensitive only to the magnitude $v_{\text{lab}} = |\mathbf{v}_{\text{lab}}|$. Second, the time dependence of $v_{\text{lab}}$ is dominated by the orbital motion of the Earth and can be neglected for short experiments (when $T \ll 1$~year). On the other hand, the direction of $\mathbf{v}_{\text{lab}}$ varies daily due to the Earth's rotation. Moreover, we show below that the gradient signal is affected more strongly by the changes in the direction of $\mathbf{v}_{\text{lab}}$ than by the annual variations in its magnitude.

The magnitude of $\mathbf{v}_{\text{lab}}$ is given, neglecting the 0.2\% contribution from the Earth's rotation, by the following expression~\cite{Knirck_2018, Foster_2018}:
\begin{equation}
v_{\text{lab}}(t) = \sqrt{v_{\odot}^2 + v_{\oplus}^2 + \eta \, v_{\odot} v_{\oplus} \cos{\left[\omega_y (t - \tau)\right]}}, \label{eq:v_lab_magnitude}
\end{equation}
where $v_{\odot} = 233~\text{km}/\text{s}$ is the speed of the Sun in the galactic rest frame, $v_{\oplus} = 29.8~\text{km} / \text{s}$ is the orbital speed of the Earth revolving around the Sun, $\omega_{y} = 2\pi / (365~\text{days})$ is the Earth's orbital angular speed, and $\eta \approx 0.982$ accounts for the inclination angle of about $60^{\circ}$ between the Earth's orbit and the galactic plane. Note that the rotational speed of the Earth at the equator is about $0.47~\text{km}/\text{s}$, which is negligible compared to~$v_{\oplus}$. The time offset in Eq.~\eqref{eq:v_lab_magnitude} is $\tau = t_y + \bar t$, where $t_y$ is the time of the vernal equinox (occurring usually on March 20) and $\bar t = 72.4$ days.

\begin{figure*}
\includegraphics[width=\textwidth]{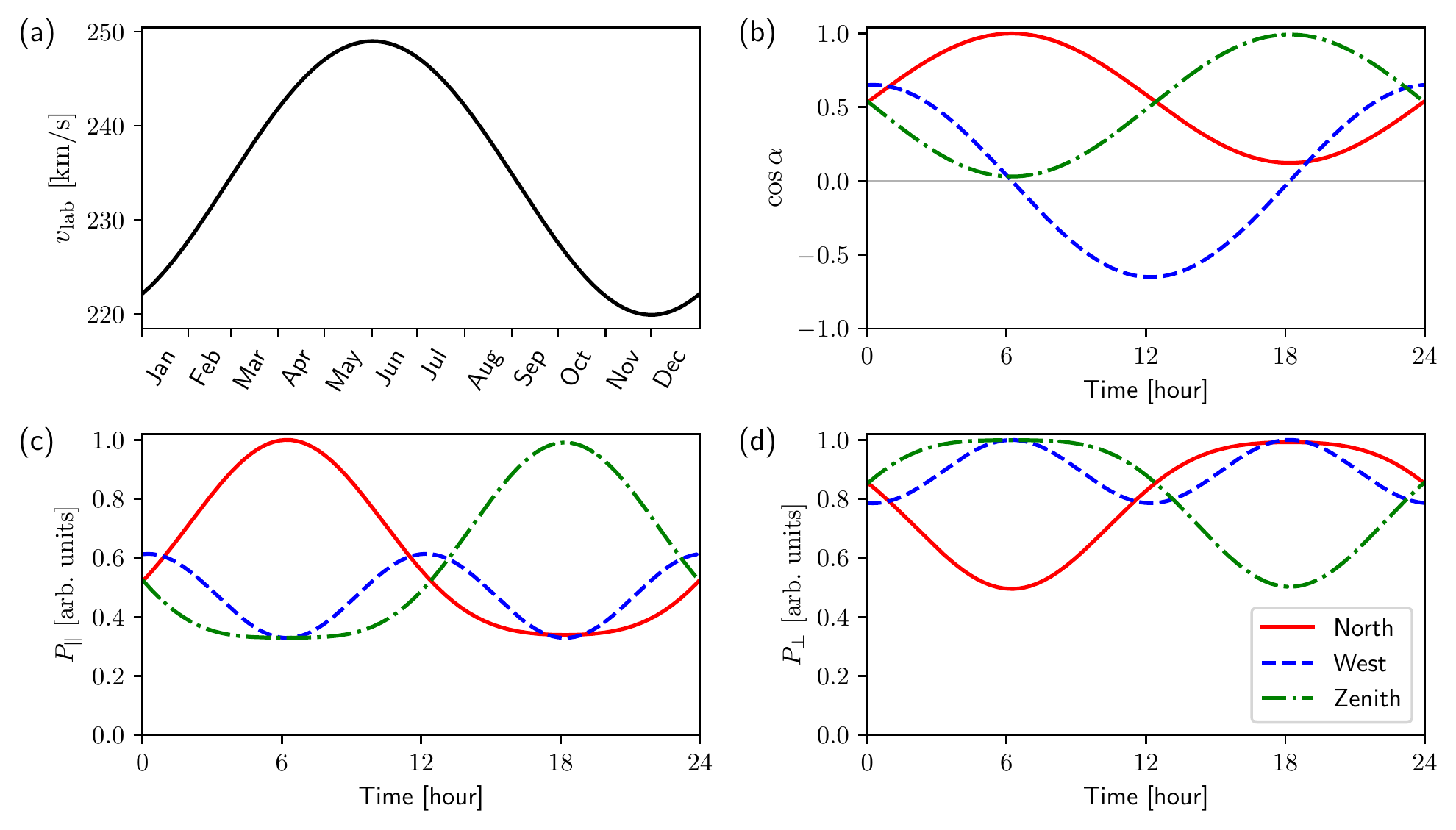}
\caption{\label{fig:modulations}Periodic modulations of $\mathbf{v}_{\text{lab}}$ and of the signal power in the case of gradient coupling. (a) Annual modulation of $v_{\text{lab}}$ due to the Earth's orbital motion around the Sun. (b)~Daily modulations of $\cos{\alpha}$ for the three orthogonal orientations of an external magnetic field: towards the north (solid red line), towards the west (dashed blue line), and towards the zenith (dash-dotted green line). (c)~Daily modulations of the total signal power $P_{\parallel}$ for the three orientations of an external magnetic field. The signal power is normalized such that the maximum value is~1 (the scaling factor is the same for the three cases). (d)~Similar to panel~(c) but for the power $P_{\perp}$. For panels (b)--(d), we assume that the location is the Metcalf Science Center of Boston University ($\lambda_{\text{lab}} = 42.3484^{\circ}$, $\phi_{\text{lab}} = -71.1002^{\circ}$) and the date is January 1 (from 00:00 to 24:00 in local time).}
\end{figure*}

The dependence $v_{\text{lab}}(t)$ given by Eq.~\eqref{eq:v_lab_magnitude} is shown in Fig.~\ref{fig:modulations}(a) for a one-year period starting on January~1. We can see that $v_{\text{lab}}$ varies by only 13\%, from $220~\text{km}/\text{s}$ (around December~1) to $249~\text{km}/\text{s}$ (around June~1). This annual variation corresponds to the 19\% and 14\% changes in the signal powers $P_{\parallel}$ and $P_{\perp}$, respectively, assuming that the angles $\alpha$ are chosen to maximize the sensitivity ($\alpha = 0$ for $P_{\parallel}$ and $\alpha = \pi/2$ for $P_{\perp}$). As we will see below, typical variations in the signal power due to daily modulations of $\alpha$ are significantly larger (of the order of 100\%).

To specify the direction of $\mathbf{v}_{\text{lab}}$, we use the coordinate system $(\hat{\mathcal{N}}, \, \hat{\mathcal{W}}, \, \hat{\mathcal{Z}})$, which is given by the three mutually orthogonal vectors pointing to the north, to the west, and to the zenith, respectively. The angles between $\mathbf{v}_{\text{lab}}$ and these three vectors can be written as~\cite{Knirck_2018}
\begin{align}
\cos{\alpha_{N}(t)} &= b_0 \cos{\lambda_{\text{lab}}} - b_1 \sin{\lambda_{\text{lab}}} \cos{\left(\omega_d t + \phi\right)}, \label{eq:alpha_N} \\
\cos{\alpha_{W}(t)} &= b_1 \sin{\left(\omega_d t + \phi\right)}, \\
\cos{\alpha_{Z}(t)} &= b_0 \sin{\lambda_{\text{lab}}} + b_1 \cos{\lambda_{\text{lab}}} \cos{\left(\omega_d t + \phi\right)}, \label{eq:alpha_Z}
\end{align}
where $\phi = \phi_{\text{lab}} + \psi$ is the phase, $\lambda_{\text{lab}}$ and $\phi_{\text{lab}}$ are the latitude and the longitude of the laboratory, and $\omega_d = 2\pi / (0.9973~\text{days})$ is the Earth's rotational angular speed. The time $t$ in Eqs.~\eqref{eq:alpha_N}--\eqref{eq:alpha_Z} is measured from the beginning of January~1. Although the parameters $b_0$, $b_1$, and $\psi$ vary during the year, they can be considered constant on a timescale of several days. The corresponding values on January~1 are $b_0 = 0.7589$, $b_1 = 0.6512$, and $\psi = -3.5336$. Full analytical expressions for $b_0$, $b_1$, and $\psi$ as functions of time are provided in Ref.~\cite{Knirck_2018}.

Figure~\ref{fig:modulations}(b) shows the time dependences \eqref{eq:alpha_N}--\eqref{eq:alpha_Z} for a 24-hour period on January~1 and for the location of the Metcalf Science Center of Boston University. Note that the cases $\alpha = \alpha_N$, $\alpha = \alpha_W$, and $\alpha = \alpha_Z$ correspond to the external magnetic field $\mathbf{B}_0$ oriented towards the north, the west, and the zenith, respectively. The daily variations in $\cos{\alpha}$ lead to corresponding modulations in the signal powers $P_{\parallel}$ and $P_{\perp}$, as shown in Figs.~\ref{fig:modulations}(c) and \ref{fig:modulations}(d). We can see that the signal power varies during the day by as much as a factor of 3. In the case of detection, these daily modulations will be a powerful confirmation that the signal is correlated with the Earth's rotation with respect to the galactic DM halo. The amplitude and phase of the modulations are deterministic and can be predicted for a specific time and location. Going a step further, one can put the haloscope on a rotating platform and modulate the signal in a controlled way.

In addition to the total signal power, the spectral line shape is also daily modulated due to its dependence on the angle $\alpha$ shown in Fig.~\ref{fig:lineshapes}(a). This dependence can serve as an additional nontrivial signature of axionlike DM. For example, one can divide the collected time-domain data in several subsets taken at the same time of day and compare the shapes of the corresponding signals in frequency domain. The most complete information can be obtained by analyzing data from three haloscopes having mutually orthogonal sensitivity axes.

\section{Conclusion}

In this paper, we have considered spectral line shapes and power spectra of the expected signal for a haloscope experiment searching for axionlike DM in our Galaxy. Assuming the standard halo model, we have rederived the spectral line shape~\eqref{eq:non-gradient_lineshape} that has been previously obtained for the nongradient couplings in Refs.~\cite{Turner_1990, O'Hare_2017, Derevianko_2018, Foster_2018}. Our derivation is straightforward and based on the connection~\eqref{eq:lambda_nu} between the line shape $\lambda(\nu)$ and the speed distribution $f(v)$ of ALPs in the galactic halo, as seen in the laboratory frame. We have extended this derivation to the gradient coupling and have considered experiments sensitive to a specific component---either parallel or perpendicular---of the ALP field gradient with respect to the direction of applied static magnetic field. The resulting spectral line shape and power spectrum are given by Eqs.~\eqref{eq:lambda_par} and \eqref{eq:S_par} in the parallel case and by Eqs.~\eqref{eq:lambda_perp} and \eqref{eq:S_perp} in the perpendicular case. To independently check our formulas, we have also performed a Monte Carlo simulation based on the stochastic model of the ALP field given by Eq.~\eqref{eq:axion_field}. The simulated power spectra agree with the analytical results, as shown in Fig.~\ref{fig:lineshapes}. Finally, we have discussed the daily and annual modulations of the signal in the case of gradient coupling. We have demonstrated in Fig.~\ref{fig:modulations} that the directional sensitivity of a gradient haloscope leads to strong daily modulations of the total signal power.

We would like to conclude by reiterating the advantages of the gradient coupling for axionlike DM searches. One can achieve directional sensitivity with a single gradient haloscope, while in the case of ALP field couplings this would require two spatially separated detectors~\cite{Foster_2021}. The directional sensitivity leads to strong daily modulations that, in the case of detection, would greatly help in confirming the DM nature of the signal. Another advantage is that the gradient coupling allows one to probe the 3D velocity distribution of ALPs in the galactic halo, thus paving the way to full-fledged ``axion astronomy''~\cite{O'Hare_2017}. Finally, since one can simultaneously probe three independent spatial directions, a gradient haloscope is less susceptible to stochastic amplitude fluctuations of the ALP field that may reduce the sensitivity of a nongradient experiment by as much as an order of magnitude~\cite{Centers_2019}. We also point out that a gradient haloscope may have enhanced sensitivity to the relativistic cosmic axion background~\cite{Dror_2021}.

\begin{acknowledgments}
The authors at Boston University acknowledge support from the Simons Foundation Grant No.\ 641332, the National Science Foundation CAREER Grant No.\ PHY-2145162, the John Templeton Foundation Grant No.\ 60049570, and the U.S. Department of Energy, Office of High Energy Physics under the QuantISED program, FWP 100495. The work of the Mainz group was supported by the Cluster of Excellence ``Precision Physics, Fundamental Interactions, and Structure of Matter'' (PRISMA+ EXC 2118/1) funded by the German Research Foundation (DFG) within the German Excellence Strategy (Project ID~39083149), by the European Research Council (ERC) under the European Union Horizon 2020 research and innovation program (project Dark-OST, Grant Agreement No.\ 695405), by the DFG Reinhart Koselleck project, and by the German Federal Ministry of Education and Research (BMBF) within the Quantumtechnologien program (Grant No.\ 13N15064). D.~F.~J.~K. acknowledges the support of the U.S. National Science Foundation under Grants No.\ PHY-1707875 and No.\ PHY-2110388. A.~W. is grateful to Professor Achim Klenke for insightful discussions on the central limit theorem.
\end{acknowledgments}

\appendix
\section{Two-dimensional random walk with a variable step size}
\label{sec:Appendix_A}

Here we provide additional information on isotropic two-dimensional random walks~\cite{Rayleigh_1880, Rayleigh_1919, Chandrasekhar_1943} as well as a derivation of Eq.~\eqref{eq:summation_cos}. As discussed in Sec.~\ref{sec:statistics}, the end point of an $N$-step walk can be described by the complex random variable
\begin{equation}
z = x + i y = \sum\limits_{n=1}^{N} s_n \exp{(i \phi_n)}, \label{eq:z_sum}
\end{equation}
where the size~$s_n$ of each step is drawn from a specific probability distribution with mean~$\mu_s$ and variance~$\sigma_s^2$, while the direction angle, $\phi_n \in [0, \, 2\pi)$, follows a uniform distribution. We assume that $s_n$ and $\phi_n$ are statistically independent, which, in combination with the central limit theorem, allows us to show that $x$ and $y$ (the real and imaginary parts of~$z$) are distributed normally.

Indeed, if $r_1$ and $r_2$ are two independent random variables with means $\mu_1$, $\mu_2$ and variances $\sigma_1^2$, $\sigma_2^2$, respectively, then the probability distribution of the product $r_1 r_2$ has the expected (mean) value
\begin{equation}
\E{(r_1 r_2)} = \mu_1 \mu_2
\end{equation}
and the variance
\begin{equation}
\Var{(r_1 r_2)} = \left(\mu_1^2 + \sigma_1^2\right) \left(\mu_2^2 + \sigma_2^2\right) - \mu_1^2 \mu_2^2.
\end{equation}
Since
\begin{equation}
x = \sum\limits_{n=1}^{N} s_n \cos{\phi_n}, \qquad y = \sum\limits_{n=1}^{N} s_n \sin{\phi_n}, \label{eq:x_y_sums}
\end{equation}
and
\begin{gather}
\E{(\cos{\phi_n})} = \E{(\sin{\phi_n})} = 0, \\
\Var{(\cos{\phi_n})} = \Var{(\sin{\phi_n})} = \frac{1}{2},
\end{gather}
we immediately conclude that
\begin{gather}
\E{(s_n \cos{\phi_n})} = \E{(s_n \sin{\phi_n})} = 0, \label{eq:product_expectations} \\
\Var{(s_n \cos{\phi_n})} = \Var{(s_n \sin{\phi_n})} = \frac{1}{2} \left(\mu_s^2 + \sigma_s^2\right). \label{eq:product_variances}
\end{gather}
We then use the central limit theorem~\cite{Papoulis&Pillai}, which states that the distribution of a sum of $N$ independent and identically distributed random variables with mean~$\mu$ and finite variance~$\sigma^2$ approaches, as $N$ increases, a normal (Gaussian) distribution with mean~$N \mu$ and variance $N \sigma^2$. After applying this theorem to the sums~\eqref{eq:x_y_sums} and taking into account Eqs.~\eqref{eq:product_expectations} and~\eqref{eq:product_variances}, we conclude that, as long as $N \gg 1$, both $x$ and $y$ follow the normal distribution with zero mean and variance $N (\mu_s^2 + \sigma_s^2) / 2$. Note that this conclusion does not rely on our knowledge of the specific probability distribution for $s_n$ (besides its mean and variance).

Thus, $z$ is a complex Gaussian random variable with zero mean and variance $\sigma^2 = N (\mu_s^2 + \sigma_s^2) / 2$. In the polar form, it can be written as $z = r \exp{(i \phi')}$, where
\begin{equation}
r = \sqrt{x^2 + y^2}, \qquad \phi' = \atan2{(y, \, x)}. \label{eq:r_phi}
\end{equation}
It is well known (see Example 6-15 in Ref.~\cite{Papoulis&Pillai}) that the magnitude $r$, which corresponds to the distance between the end point of the walk and the origin, follows the Rayleigh distribution
\begin{equation}
p(r; \, \sigma) = \frac{r}{\sigma^2} \exp{\left(-\frac{r^2}{2\sigma^2}\right)}. \label{eq:Rayleigh_appendix}
\end{equation}
The argument $\phi'$ is distributed uniformly, which reflects the fact that there is no preferred direction for an isotropic random walk. The squared distance, $r^2$, follows the exponential probability distribution
\begin{equation}
p(r^2; \, \sigma) = \frac{1}{2\sigma^2} \exp{\left(-\frac{r^2}{2\sigma^2}\right)},
\end{equation}
as demonstrated in Example 6-14 of Ref.~\cite{Papoulis&Pillai}. This distribution appears in Eq.~\eqref{eq:exponential_distribution} and describes the PSD values sampled at each frequency~$\nu_j$.

Let us now show how to derive Eq.~\eqref{eq:summation_cos}. The left-hand side of this equation can be rewritten as
\begin{gather}
\sum\limits_{n=1}^{N} s_n \cos{(\omega t + \phi_n)} = \re{\left(e^{i \omega t} \sum\limits_{n=1}^{N} s_n e^{i \phi_n}\right)} \nonumber \\
= \re{\left(e^{i \omega t} z\right)} = r \cos{(\omega t + \phi')}, \label{eq:sum_appendix}
\end{gather}
where the angular frequency, $\omega = 2\pi \nu$, is used for brevity. Recall that $r$ follows the Rayleigh distribution~\eqref{eq:Rayleigh_appendix} with variance $\sigma^2 = N (\mu_s^2 + \sigma_s^2) / 2$. Finally, we introduce a normalized variable, $r' = r / \sigma$, and the right-hand side of Eq.~\eqref{eq:sum_appendix} becomes
\begin{equation}
r \cos{(\omega t + \phi')} = \sqrt{\frac{N (\mu_s^2 + \sigma_s^2)}{2}} \, r' \cos{(\omega t + \phi')},
\end{equation}
where $r'$ follows the Rayleigh distribution~\eqref{eq:Rayleigh_appendix} with unit variance.

\section{Expressions for $\mu_{\parallel, \perp}$ and $\sigma_{\parallel, \perp}^2$\\ and a derivation of Eq.~\eqref{eq:deterministic_factor}}
\label{sec:Appendix_B}

The mean $\mu_{\parallel, \perp}$ and the variance $\sigma_{\parallel, \perp}^2$ introduced in Eq.~\eqref{eq:nabla_a_stochastic} are defined as
\begin{gather}
\mu_{\parallel, \perp}(v) = \frac{v^2}{f_{\text{lab}}(v)} \iint v_{\parallel, \perp} f_{\text{lab}}^{(3)}(\mathbf{v}) \, d\Omega, \label{eq:mean} \\
\sigma_{\parallel, \perp}^2(v) = \frac{v^2}{f_{\text{lab}}(v)} \iint \left(v_{\parallel, \perp} - \mu_{\parallel, \perp}\right)^2 f_{\text{lab}}^{(3)}(\mathbf{v}) \, d\Omega, \label{eq:variance}
\end{gather}
where for brevity we have denoted the integral over the solid angle as
\begin{equation}
\int\limits_{0}^{2\pi} d\phi \int\limits_{0}^{\pi} \ldots \, \sin{\theta} \, d\theta = \iint \ldots \, d\Omega.
\end{equation}
Let us show how Eq.~\eqref{eq:deterministic_factor} can be derived assuming these definitions. We first note that
\begin{equation}
\frac{v^2}{f_{\text{lab}}(v)} \iint f_{\text{lab}}^{(3)}(\mathbf{v}) \, d\Omega = 1, \label{eq:unit_integral}
\end{equation}
as immediately follows from Eq.~\eqref{eq:f_lab_integral}. Then, we rewrite Eq.~\eqref{eq:variance} as
\begin{align}
\sigma_{\parallel, \perp}^2 &= \frac{v^2}{f_{\text{lab}}(v)} \iint v_{\parallel, \perp}^2 f_{\text{lab}}^{(3)}(\mathbf{v}) \, d\Omega \nonumber \\
{} &- 2\mu_{\parallel, \perp} \frac{v^2}{f_{\text{lab}}(v)} \iint v_{\parallel, \perp} f_{\text{lab}}^{(3)}(\mathbf{v}) \, d\Omega \nonumber \\
{} &+ \mu_{\parallel, \perp}^2 \frac{v^2}{f_{\text{lab}}(v)} \iint f_{\text{lab}}^{(3)}(\mathbf{v}) \, d\Omega \nonumber \\
&= \frac{v^2}{f_{\text{lab}}(v)} \iint v_{\parallel, \perp}^2 f_{\text{lab}}^{(3)}(\mathbf{v}) \, d\Omega - \mu_{\parallel, \perp}^2,
\end{align}
where at the last step we used Eqs.~\eqref{eq:mean} and~\eqref{eq:unit_integral}. Therefore, we have shown that
\begin{gather}
\bigl(\mu_{\parallel, \perp}^2 + \sigma_{\parallel, \perp}^2\bigr) f_{\text{lab}}(v) = v^2 \iint v_{\parallel, \perp}^2 f_{\text{lab}}^{(3)}(\mathbf{v}) \, d\Omega \nonumber \\
= C_{\parallel, \perp} f_{\parallel, \perp}(v),
\end{gather}
where the last step follows from Eq.~\eqref{eq:f_lab_integration}.


\begin{thebibliography}{99}
\bibitem{Bertone_2005}
G.~Bertone, D.~Hooper, and J.~Silk, Particle dark matter: Evidence, candidates and constraints, \href{https://doi.org/10.1016/j.physrep.2004.08.031}{Phys. Rep. \textbf{405}, 279 (2005)}.

\bibitem{Feng_2010}
J.~L.~Feng, Dark matter candidates from particle physics and methods of detection, \href{https://doi.org/10.1146/annurev-astro-082708-101659}{Annu. Rev. Astron. Astrophys. \textbf{48}, 495 (2010)}.

\bibitem{PDG_2020}
P.~A.~Zyla \textit{et al.} (Particle Data Group), Review of particle physics, \href{https://doi.org/10.1093/ptep/ptaa104}{Prog. Theor. Exp. Phys. (\textbf{2020}), 083C01}.

\bibitem{White_1978}
S.~D.~M.~White and M.~J.~Rees, Core condensation in heavy halos: A two-stage theory for galaxy formation and clustering, \href{https://doi.org/10.1093/mnras/183.3.341}{Mon. Not. R. Astron. Soc. \textbf{183}, 341 (1978)}.

\bibitem{Wechsler_2018}
R.~H.~Wechsler and J.~L.~Tinker, The connection between galaxies and their dark matter halos, \href{https://doi.org/10.1146/annurev-astro-081817-051756}{Annu. Rev. Astron. Astrophys. \textbf{56}, 435 (2018)}.

\bibitem{Graham_2015}
P.~W.~Graham, I.~G.~Irastorza, S.~K.~Lamoreaux, A.~Lindner, and K.~A.~van~Bibber, Experimental searches for the axion and axion-like particles, \href{https://doi.org/10.1146/annurev-nucl-102014-022120}{Annu. Rev. Nucl. Part. Sci. \textbf{65}, 485 (2015)}.

\bibitem{Irastorza_2018}
I.~G.~Irastorza and J.~Redondo, New experimental approaches in the search for axion-like particles, \href{https://doi.org/10.1016/j.ppnp.2018.05.003}{Prog. Part. Nucl. Phys. \textbf{102}, 89 (2018)}.

\bibitem{Sikivie_2021}
P.~Sikivie, Invisible axion search methods, \href{https://doi.org/10.1103/RevModPhys.93.015004}{Rev. Mod. Phys. \textbf{93}, 015004 (2021)}.

\bibitem{Graham_2013}
P.~W.~Graham and S.~Rajendran, New observables for direct detection of axion dark matter, \href{https://doi.org/10.1103/PhysRevD.88.035023}{Phys. Rev. D \textbf{88}, 035023 (2013)}.

\bibitem{Sikivie_1983}
P.~Sikivie, Experimental tests of the ``invisible'' axion, \href{https://doi.org/10.1103/PhysRevLett.51.1415}{Phys. Rev. Lett. \textbf{51}, 1415 (1983)}.

\bibitem{Krauss_1985}
L.~Krauss, J.~Moody, F.~Wilczek, and D.~E.~Morris, Calculations for cosmic axion detection, \href{https://doi.org/10.1103/PhysRevLett.55.1797}{Phys. Rev. Lett. \textbf{55}, 1797 (1985)}.

\bibitem{Turner_1990}
M.~S.~Turner, Periodic signatures for the detection of cosmic axions, \href{https://doi.org/10.1103/PhysRevD.42.3572}{Phys. Rev. D \textbf{42}, 3572 (1990)}.

\bibitem{Du_2018}
N.~Du \textit{et al.} (ADMX Collaboration), Search for invisible axion dark matter with the Axion Dark Matter Experiment, \href{https://doi.org/10.1103/PhysRevLett.120.151301}{Phys. Rev. Lett. \textbf{120}, 151301 (2018)}.

\bibitem{Braine_2020}
T.~Braine \textit{et al.} (ADMX Collaboration), Extended search for the invisible axion with the Axion Dark Matter Experiment, \href{https://doi.org/10.1103/PhysRevLett.124.101303}{Phys. Rev. Lett. \textbf{124}, 101303 (2020)}.

\bibitem{Lee_2020}
S.~Lee, S.~Ahn, J.~Choi, B.~R.~Ko, and Y.~K.~Semertzidis, Axion dark matter search around $6.7~\mu\text{eV}$, \href{https://doi.org/10.1103/PhysRevLett.124.101802}{Phys. Rev. Lett. \textbf{124}, 101802 (2020)}.

\bibitem{Kwon_2021}
O.~Kwon \textit{et al.}, First results from an axion haloscope at CAPP around $10.7~\mu\text{eV}$, \href{https://doi.org/10.1103/PhysRevLett.126.191802}{Phys. Rev. Lett. \textbf{126}, 191802 (2021)}.

\bibitem{Brubaker_PRD_2017}
B.~M. Brubaker, L. Zhong, S.~K. Lamoreaux, K.~W. Lehnert, and K.~A. van~Bibber, HAYSTAC axion search analysis procedure, \href{https://doi.org/10.1103/PhysRevD.96.123008}{Phys. Rev. D \textbf{96}, 123008 (2017)}.

\bibitem{Brubaker_PRL_2017}
B.~M.~Brubaker \textit{et al.}, First results from a microwave cavity axion search at $24~\mu\text{eV}$, \href{https://doi.org/10.1103/PhysRevLett.118.061302}{Phys. Rev. Lett. \textbf{118}, 061302 (2017)}.

\bibitem{Backes_2021}
K.~M.~Backes \textit{et al.}, A quantum enhanced search for dark matter axions, \href{https://doi.org/10.1038/s41586-021-03226-7}{Nature (London) \textbf{590}, 238 (2021)}.

\bibitem{Gramolin_2021}
A.~V.~Gramolin, D.~Aybas, D.~Johnson, J.~Adam, and A.~O.~Sushkov, Search for axion-like dark matter with ferromagnets, \href{https://doi.org/10.1038/s41567-020-1006-6}{Nat. Phys. \textbf{17}, 79 (2021)}.

\bibitem{Foster_2018}
J.~W.~Foster, N.~L.~Rodd, and B.~R.~Safdi, Revealing the dark matter halo with axion direct detection, \href{https://doi.org/10.1103/PhysRevD.97.123006}{Phys. Rev. D \textbf{97}, 123006 (2018)}.

\bibitem{Loudon}
R.~Loudon, \textit{The Quantum Theory of Light} (Oxford University Press, New York, 2000).

\bibitem{Budker_2014}
D.~Budker, P.~W.~Graham, M.~Ledbetter, S.~Rajendran, and A.~O.~Sushkov, Proposal for a Cosmic Axion Spin Precession Experiment (CASPEr), \href{https://doi.org/10.1103/PhysRevX.4.021030}{Phys. Rev. X \textbf{4}, 021030 (2014)}.

\bibitem{Abel_2017}
C.~Abel \textit{et al.}, Search for axionlike dark matter through nuclear spin precession in electric and magnetic fields, \href{https://doi.org/10.1103/PhysRevX.7.041034}{Phys. Rev. X \textbf{7}, 041034 (2017)}.

\bibitem{Terrano_2019}
W.~A.~Terrano, E.~G.~Adelberger, C.~A.~Hagedorn, and B.~R.~Heckel, Constraints on axionlike dark matter with masses down to $10^{-23}~\text{eV}/c^2$, \href{https://doi.org/10.1103/PhysRevLett.122.231301}{Phys. Rev. Lett. \textbf{122}, 231301 (2019)}.

\bibitem{Garcon_2019}
A.~Garcon \textit{et al.}, Constraints on bosonic dark matter from ultralow-field nuclear magnetic resonance, \href{https://doi.org/10.1126/sciadv.aax4539}{Sci. Adv. \textbf{5}, eaax4539 (2019)}.

\bibitem{Wu_2019}
T.~Wu \textit{et al.}, Search for axionlike dark matter with a liquid-state nuclear spin comagnetometer, \href{https://doi.org/10.1103/PhysRevLett.122.191302}{Phys. Rev. Lett. \textbf{122}, 191302 (2019)}.

\bibitem{Jiang_2021}
M.~Jiang, H.~Su, A.~Garcon, X.~Peng, and D.~Budker, Search for axion-like dark matter with spin-based amplifiers, \href{https://doi.org/10.1038/s41567-021-01392-z}{Nat. Phys. \textbf{17}, 1402 (2021)}.

\bibitem{Bloch_2021}
I.~M.~Bloch, G.~Ronen, R.~Shaham, O.~Katz, T.~Volansky, and O.~Katz, New constraints on axion-like dark matter using a Floquet quantum detector, \href{https://doi.org/10.1126/sciadv.abl8919}{Sci. Adv. \textbf{8}, eabl8919 (2022)}. 

\bibitem{Centers_2019}
G.~P.~Centers \textit{et al.}, Stochastic fluctuations of bosonic dark matter, \href{https://doi.org/10.1038/s41467-021-27632-7}{Nat. Commun. \textbf{12}, 7321 (2021)}.

\bibitem{Lisanti_2021}
M.~Lisanti, M.~Moschella, and W.~Terrano, Stochastic properties of ultralight scalar field gradients, \href{https://doi.org/10.1103/PhysRevD.104.055037}{Phys. Rev. D \textbf{104}, 055037 (2021)}.

\bibitem{Aybas_2021}
D.~Aybas \textit{et al.}, Search for axionlike dark matter using solid-state nuclear magnetic resonance, \href{https://doi.org/10.1103/PhysRevLett.126.141802}{Phys. Rev. Lett. \textbf{126}, 141802 (2021)}.

\bibitem{Crescini_2018}
N.~Crescini \textit{et al.}, Operation of a ferromagnetic axion haloscope at $m_a = 58~\mu\text{eV}$, \href{https://doi.org/10.1140/epjc/s10052-018-6163-8}{Eur. Phys. J. C \textbf{78}, 703 (2018)}.

\bibitem{Crescini_2020}
N.~Crescini \textit{et al.} (QUAX Collaboration), Axion search with a quantum-limited ferromagnetic haloscope, \href{https://doi.org/10.1103/PhysRevLett.124.171801}{Phys. Rev. Lett. \textbf{124}, 171801 (2020)}.

\bibitem{Derevianko_2018}
A.~Derevianko, Detecting dark-matter waves with a network of precision-measurement tools, \href{https://doi.org/10.1103/PhysRevA.97.042506}{Phys. Rev. A \textbf{97}, 042506 (2018)}.

\bibitem{Foster_2021}
J.~W.~Foster, Y.~Kahn, R.~Nguyen, N.~L.~Rodd, and B.~R.~Safdi, Dark matter interferometry, \href{https://doi.org/10.1103/PhysRevD.103.076018}{Phys. Rev. D \textbf{103}, 076018 (2021)}.

\bibitem{Schumann_2019}
M.~Schumann, Direct detection of WIMP dark matter: Concepts and status, \href{https://doi.org/10.1088/1361-6471/ab2ea5}{J. Phys. G \textbf{46}, 103003 (2019)}.

\bibitem{Evans_2019}
N.~W.~Evans, C.~A.~J.~O’Hare, and C.~McCabe, Refinement of the standard halo model for dark matter searches in light of the Gaia Sausage, \href{https://doi.org/10.1103/PhysRevD.99.023012}{Phys. Rev. D \textbf{99}, 023012 (2019)}.

\bibitem{O'Hare_2017}
C.~A.~J.~O’Hare and A.~M.~Green, Axion astronomy with microwave cavity experiments, \href{https://doi.org/10.1103/PhysRevD.95.063017}{Phys. Rev. D \textbf{95}, 063017 (2017)}.

\bibitem{GitHub}
\url{https://github.com/gramolin/lineshape}.

\bibitem{Drukier_1986}
A.~K.~Drukier, K.~Freese, and D.~N.~Spergel, Detecting cold dark-matter candidates, \href{https://doi.org/10.1103/PhysRevD.33.3495}{Phys. Rev. D \textbf{33}, 3495 (1986)}.

\bibitem{Freese_2013}
K.~Freese, M.~Lisanti, and C.~Savage, Colloquium: Annual modulation of dark matter, \href{https://doi.org/10.1103/RevModPhys.85.1561}{Rev. Mod. Phys. \textbf{85}, 1561 (2013)}.

\bibitem{Knirck_2018}
S.~Knirck, A.~J.~Millar, C.~A.~J.~O’Hare, J.~Redondo, and F.~D.~Steffen, Directional axion detection, \href{https://doi.org/10.1088/1475-7516/2018/11/051}{J. Cosmol. Astropart. Phys. 11 (2018) 051}.

\bibitem{Dror_2021}
J.~A.~Dror, H.~Murayama, and N.~L.~Rodd, Cosmic axion background, \href{https://doi.org/10.1103/PhysRevD.103.115004}{Phys. Rev. D \textbf{103}, 115004 (2021)}.

\bibitem{Rayleigh_1880}
Lord Rayleigh, On the resultant of a large number of vibrations of the same pitch and of arbitrary phase, \href{https://doi.org/10.1080/14786448008626893}{Philos. Mag. \textbf{10}, 73 (1880)}.

\bibitem{Rayleigh_1919}
Lord Rayleigh, On the problem of random vibrations, and of random flights in one, two, or three dimensions, \href{https://doi.org/10.1080/14786440408635894}{Philos. Mag. \textbf{37}, 321 (1919)}.

\bibitem{Chandrasekhar_1943}
S.~Chandrasekhar, Stochastic problems in physics and astronomy, \href{https://doi.org/10.1103/RevModPhys.15.1}{Rev. Mod. Phys. \textbf{15}, 1 (1943)}.

\bibitem{Papoulis&Pillai}
A.~Papoulis and S.~U.~Pillai, \textit{Probability, Random Variables, and Stochastic Processes} (McGraw-Hill, New York, 2002).
\end{thebibliography}
\end{document}